\documentclass[letterpaper]{article} 
\usepackage{aaai25}
\usepackage{times}  
\usepackage{helvet}  
\usepackage{courier}  
\usepackage[hyphens]{url}  
\usepackage{graphicx} 
\usepackage{natbib}  
\usepackage{caption} 
\usepackage{algorithm}
\usepackage{algorithmic}
\usepackage{tikz-cd}
\usetikzlibrary{arrows}

\usepackage{tikz}
\usepackage{tikz-qtree,tikz-qtree-compat}
\usetikzlibrary{positioning}
\usepackage{amsmath}
\usepackage{subcaption}
\usepackage{graphicx}
\usepackage{multirow}

\usepackage{newfloat}
\usepackage{listings}
\DeclareCaptionStyle{ruled}{labelfont=normalfont,labelsep=colon,strut=off} 
\floatstyle{ruled}
\newfloat{listing}{tb}{lst}{}
\floatname{listing}{Listing}
\title{Causality in the Can: Diet Coke's Impact on Fatness}
\author{
    Yicheng Qi\textsuperscript{\rm 1},
    Ang Li\textsuperscript{\rm 2}
}
\affiliations{
    \textsuperscript{\rm 1}Soochow University, Suzhou, China\\
    \textsuperscript{\rm 2}Florida State University, Tallahassee, FL, USA\\

    ycqi@stu.suda.edu.cn, angli@cs.fsu.edu
}

\usepackage{bibentry}

\begin{document}

\maketitle

\begin{abstract}
Artificially sweetened beverages like Diet Coke are often considered better alternatives to sugary drinks, but the debate over their impact on health, particularly in relation to obesity, continues. Previous research has predominantly used association-based methods with observational or Randomized Controlled Trial (RCT) data\footnote{Randomized Controlled Trial (RCT), also known as an A/B test, is a type of scientific experiment commonly used in fields such as medicine, psychology, and social sciences. In an RCT, participants are randomly assigned to one of two or more groups: the treatment group, which receives the intervention being tested, and the control group, which either receives a placebo or no treatment at all. The data collected from an RCT is referred to as RCT/experimental data. In contrast, data collected from a survey or study where participants select their own treatment is known as observational data.}, which may not accurately capture the causal relationship between Diet Coke consumption and obesity, leading to potentially limited conclusions. In contrast, we employed causal inference methods using Structural Causal Models, integrating both observational and RCT data. Specifically, we utilized data from the National Health and Nutrition Examination Survey (NHANES), which includes diverse demographic information, as our observational data source. This data was then used to construct a causal graph, and the back-door criterion, along with its adjustment formula, was applied to estimate the RCT data. We then calculated the counterfactual quantity, the Probability of Necessity and Sufficiency (PNS), using both NHANES data and estimated RCT data. We propose that PNS is the essential metric for assessing the impact of Diet Coke on obesity. Our results indicate that between $20\%$ to $50\%$ of individuals, especially those with poor dietary habits, are more likely to gain weight from Diet Coke. Conversely, in groups like young females with healthier diets, only a small proportion experience weight gain due to Diet Coke. These findings highlight the influence of individual lifestyle and potential hormonal factors on the varied effects of Diet Coke, providing a new framework for understanding its nutritional impacts on public health.
\end{abstract}

%

\section{Introduction}

As global health consciousness rises, low-calorie artificial sweetener beverages like Diet Coke have become popular choices in the beverage market. Since its launch in 1982~\citep{spren_diet_coke_history}, Diet Coke has quickly captivated a large consumer base with its zero-sugar or low-calorie profile. Statistics~\citep{mashed_diet_coke_truth} show that Diet Coke holds a significant market share in the soft drink markets of many countries, particularly among those pursuing healthier lifestyles, where it has become the preferred daily beverage. Consumers primarily choose Diet Coke for its low-calorie properties and its potential benefits in weight management. Many people consume it as a substitute for high-sugar drinks, aiming to reduce sugar intake and control weight.

\subsection{Innovation and Contribution}

Previous studies exploring the relationship between diet soda and obesity have various shortcomings. \cite{ruanpeng2017} conducted a systematic review and meta-analysis that found a significant association between artificially sweetened beverages and obesity but did not fully account for potential confounders. \cite{Peters2014} suggested a positive role of diet beverages in weight loss but lacked detailed analysis across different demographic groups. \cite{Felstead2023} compared diet soda and water for weight loss, finding benefits for diet soda, but did not consider individual differences. \cite{Manaker2024} claimed diet soda does not lead to obesity and may help maintain a healthy weight but lacked control for confounding variables.

These studies are all association-based, relying either solely on RCTs or observational data, and they fail to adequately account for confounders. To overcome these limitations and better analyze the causal relationship between diet soda consumption and obesity, this paper employs advanced causal inference techniques:
\begin{enumerate}
    \item \textbf{Structured Causal Model (SCM)}: Using the IC* algorithm to construct a causal diagram that illustrates both direct and indirect relationships between diet soda consumption and obesity, considering potential confounders \cite{pearl1995, spirtes2000, pearl2009, koller2009}.
    \item \textbf{Back-Door Criterion Control}: Identifying and controlling for back-door paths in the causal diagram to estimate the RCT distribution of diet soda's effect using the back-door criterion and adjustment formula \cite{pearl1995}.
    \item \textbf{Probability of Necessity and Sufficiency (PNS)}: By employing the probabilities of causation and their bounds as the decision criterion \cite{TianPearl2000}, and refining these bounds with covariates, we assess the effectiveness of diet soda on obesity rates across different population groups \cite{ijcai2022p376}.
\end{enumerate}

Our results indicate that, overall, diet soda consumption may promote obesity, with significant variations across different demographic groups. For instance, individuals with poor dietary habits show a more pronounced weight gain effect from diet soda.

\subsection{Advantages}

\begin{enumerate}
    \item \textbf{Methodological comprehensiveness}: By employing advanced causal inference methods and integrating observational data with RCT estimates, this study overcomes biases inherent in single data sources, accurately accounts for the nature of counterfactual behavior in individuals, and thus reaches robust conclusions.
    \item \textbf{Reliability of NHANES Data}: This study utilizes NHANES data, known for its high representativeness and reliability due to strict multistage sampling and comprehensive data collection, with adjustments for sample weights, non-response, and population control \cite{nhanes_cdc}.
    \item \textbf{Detailed Confounder Control}: Through causal diagram construction, multiple potential confounders such as age, gender, race, education, smoking status, physical activity, hyperlipidemia, and diabetes are accurately identified and controlled, surpassing the variable control scope of existing studies.
\end{enumerate}



Through this paper, we aim to provide readers with a comprehensive perspective on the complex relationship between Diet Coke consumption and obesity, as well as the potential implications of these findings for health policies and individual choices.

\section{Preliminaries \& Related Works}

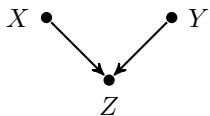
\begin{figure}[h]
\centering
\begin{tikzpicture}[->,>=stealth',node distance=1.2cm,
  thick,main node/.style={circle,fill,inner sep=1.5pt}]
  \node[main node] (1) [label=below:{$Z$}]{};
  \node[main node] (2) [above left =1cm of 1,label=left:$X$]{};
  \node[main node] (3) [above right =1cm of 1,label=right:$Y$] {};
  \path[every node/.style={font=\sffamily\small}]
    (2) edge node {} (1)
    (3) edge node {} (1);
\end{tikzpicture}
\caption{The graphical model of $SCM$, with X indicating years of schooling, Y indicating years of employment, and Z indicating salary.}
\label{causal1}
\hfill

\end{figure}

In this section, we review some fundamental methods of causal inference. First, we introduce the concept of the structural causal model, or $SCM$, as outlined in~\citep{pearl1995,spirtes2000,pearl2009,koller2009}. Figure~\ref{causal1} exemplifies an appropriate causal diagram, which is essentially a directed acyclic graph (DAG).

Formally, a $SCM$ consists of two sets of variables U and V, and a set of functions f that assigns each variable in V a value based on the values of the other variables in the model. In $SCM$, exogenous variables (U) are external and have no ancestors, depicted as root nodes in graphs. Endogenous variables (V) depend on exogenous ones and can be predicted using functions in f if all exogenous values are known.

In $SCM$s, the associated graphical model consists of nodes for each variable in U and V and directed edges that represent functional dependencies. If a variable X depends on Y, there is a directed edge from Y to X. These graphical models are typically directed acyclic graphs (DAGs). Causally, if Y is a parent of X in the graph, Y is a direct cause of X; if Y is an ancestor, it is a potential cause of X.

For instance, consider the following simple $SCM$:
\[
  U = \{X, Y\} ,  V = \{Z\} ,   F = \{f_Z\}  
\]
\[ f_Z : Z = 2X + 3Y \]

This model represents the salary (\(Z\)) that an employer pays an individual with \(X\) years of schooling and \(Y\) years in the profession. \(X\) and \(Y\) both appear in \(f_Z\), so \(X\) and \(Y\) are both direct causes of \(Z\). If \(X\) and \(Y\) had any ancestors, those ancestors would be potential causes of \(Z\).

The graphical model associated with it is illustrated in Figure 1.

One key concept of a causal diagram is called $d$-separation \citep{pearl2016causal}.

\newtheorem{definition}{Definition} 
\begin{definition}[$d$-separation]

In a causal diagram $G$, a path $p$ is blocked by a set of nodes $Z$ if and only if 
\begin{enumerate}
    \item $p$ contains a chain of nodes $A\rightarrow{}B\rightarrow{}C$ or a fork $A\leftarrow{}B\rightarrow{}C$ such that the middle node $B$ is in $Z$ (i.e., $B$ is conditioned on), or
    \item $p$ contains a collider $A\rightarrow{}B\leftarrow{}C$ such that the collision node $B$ is not in $Z$, and no descendant of $B$ is in $Z$.
\end{enumerate}
If $Z$ blocks every path between two nodes $X$ and $Y$, then $X$ and $Y$ are $d$-separated conditional on $Z$, and thus are independent conditional on $Z$, denoted as $X \perp\!\!\!\perp Y \mid Z$.
\end{definition}

With the concept of $d$-separation in a causal diagram, Pearl proposed the back-door criteria and its associated adjustment formula \cite{pearl1995} as follows:
\vspace{10pt}
\begin{definition}[Back-Door Criterion]

Given an ordered pair of variables $(X,Y)$ in a directed acyclic graph $G$, a set of variables $Z$ satisfies the back-door criterion relative to $(X,Y)$, if no node in $Z$ is a descendant of $X$, and $Z$ blocks every path between $X$ and $Y$ that contains an arrow into $X$.
\end{definition}
If a set of variables $Z$ satisfies the back-door criterion for $X$ and $Y$, the causal effects of $X$ on $Y$ are given by the adjustment formula:
\begin{eqnarray}
P(y|do(x)) = \sum_z P(y|x,z)P(z).
\label{adj1}
\end{eqnarray}

Next, we review the definitions for the three aspects
of causation as defined in ~\citep{Pearl1999}. We use the causal
diagrams ~\citep{pearl1995,spirtes2000,pearl2009,koller2009} and the language of counterfactuals in its structural model semantics, as given in ~\citep{balke2013counterfactuals,Pearl1999,Halpern2000}.

We use $Y_x=y$ to denote the counterfactual sentence ``Variable $Y$ would have the value $y$, had $X$ been $x$". For simplicity purposes, in the rest of the paper, we use $y_x$ to denote the event $Y_x=y$, $y_{x'}$ to denote the event $Y_{x'}=y$, $y'_x$ to denote the event $Y_x=y'$, and $y'_{x'}$ to denote the event $Y_{x'}=y'$. For notational simplicity, we limit the discussion to binary $X$ and $Y$, extension to multi-valued variables are straightforward \citep{pearl2009}.
\vspace{10pt}
\begin{definition}[Probability of necessity and sufficiency (PNS)] \citep{Pearl1999}
\begin{eqnarray}
\text{PNS} = P(y_x, y'_{x'})
\label{pns}
\end{eqnarray}
\end{definition}
\par
PNS stands for the probability that $y$ would respond to $x$ both ways, and therefore measures both the sufficiency and necessity of $x$ to produce $y$.

Tian and Pearl \citep{TianPearl2000} provide a tight bound for PNS without a causal diagram. Li and Pearl \citep{ijcai2019p248} provide a theoretical proof of the tight bound for PNS, and other probabilities of causation without a causal diagram.

PNS has the following tight bounds:\\

\begin{eqnarray}
\max \left \{
\begin{array}{cc}
0 \\
P(y_x) - P(y_{x'}) \\
P(y) - P(y_{x'}) \\
P(y_x) - P(y)\\
\end{array}
\right \}
\le \text{PNS}
\label{pnslb}
\end{eqnarray}

\begin{eqnarray}
\text{PNS} \le \min \left \{
\begin{array}{cc}
 P(y_x) \\
 P(y'_{x'}) \\
P(x,y) + P(x',y') \\
P(y_x) - P(y_{x'}) +\\
+ P(x, y') + P(x', y)
\end{array} 
\right \}
\label{pnsub}
\end{eqnarray}

\vspace{10pt}
Theorems \ref{thm1} and \ref{thm2} below provide bounds for PNS when a set $Z$ of variables can be measured which satisfy only one simple condition: $Z$ contains no descendant of $X$. 

\newtheorem{theorem}{Theorem}
\begin{theorem}
Given a causal diagram $G$ and distribution compatible with $G$, let $Z$ be a set of variables that does not contain any descendant of $X$ in $G$, then PNS is bounded as follows:
\begin{flushleft}
\begin{eqnarray}
\sum_z\max\left\{
\begin{array}{c}
0,\\
P(y_x|z)-P(y_{x'}|z),\\
P(y|z)-P(y_{x'}|z),\\
P(y_x|z)-P(y|z)
\end{array}
\right\}\times P(z)\le \text{PNS}
\label{inequ11}
\end{eqnarray}
\end{flushleft}
\begin{flushleft}
\begin{equation}
\sum_z\min\left\{
\begin{array}{c}
P(y_x|z),\\
P(y'_{x'}|z),\\
P(y,x|z)+P(y',x'|z),\\
P(y_x|z)-P(y_{x'}|z)+\\
+P(y,x'|z)+P(y',x|z)
\end{array}
\right\}\times P(z)\ge \text{PNS}
\label{inequ22}
\end{equation}
\end{flushleft}
\label{thm1}

\end{theorem}

\begin{figure}
\centering
\begin{subfigure}[4]{0.22\textwidth}
\centering
\begin{tikzpicture}[->,>=stealth',node distance=1cm,
  thick,main node/.style={circle,fill,inner sep=1.5pt}]
  \node[main node] (1) [label=above:{$Z$}]{};
  \node[main node] (2) [below left =1cm of 1,label=left:$X$]{};
  \node[main node] (3) [below right =1cm of 1,label=right:$Y$] {};
  \path[every node/.style={font=\sffamily\small}]
    (1) edge node {} (2)
    (1) edge node {} (3)
    (2) edge node {} (3);
\end{tikzpicture}
\caption{Confounder $Z$}
\label{causalg2}
\end{subfigure}
\begin{subfigure}[4]{0.22\textwidth}
\centering
\begin{tikzpicture}[->,>=stealth',node distance=1cm,
  thick,main node/.style={circle,fill,inner sep=1.5pt}]
  \node[main node] (1) [label=above:{$Z$}]{};
  \node[main node] (2) [below left =1cm of 1,label=left:$X$]{};
  \node[main node] (3) [below right =1cm of 1,label=right:$Y$] {};
  \path[every node/.style={font=\sffamily\small}]
    (1) edge node {} (3)
    (2) edge node {} (3);
\end{tikzpicture}
\caption{Outcome-affecting Cov $Z$}
\label{causalg3}
\end{subfigure}
\caption{$Z$ is not a descendant of $X$}
\label{causalcovariate}
\end{figure}
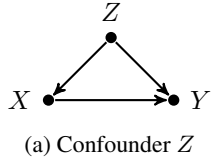
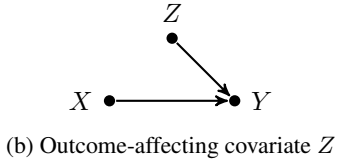

In figures \ref{causalg2} and \ref{causalg3}, $Z$ is not a descendant of $X$ and further satisfies the back-door criterion. For such cases the PNS bounds can be simplified to read:
\vspace{10pt}
\begin{theorem}
Given a causal diagram $G$ and distribution compatible with $G$, let $Z$ be a set of variables satisfying the back-door criterion \cite{pearl2011aspects} in $G$, then the PNS is bounded as follows:
\begin{flushleft}
\begin{eqnarray}
\sum_z \max\{0,P(y|x,z)-P(y|x',z)\}\times P(z) \le \text{PNS}
\label{lb_confounder}
\end{eqnarray}
\end{flushleft}
\begin{flushleft}
\begin{eqnarray}
\sum_z \min\{P(y|x,z),P(y'|x',z)\}\times P(z)
\ge \text{PNS}
\label{ub_confounder}
\end{eqnarray}
\end{flushleft}

\label{thm2}
\end{theorem}
The significance of Theorem \ref{thm2} lies in the ability to compute bounds using purely observational data, which is the situation we will face in this paper.

\section{Methodology}
\subsection{Data Collection}


All data in this paper is sourced from the NHANES program~\citep{NHANESAbout}. The National Health and Nutrition Examination Survey (NHANES), run by the CDC's National Center for Health Statistics (NCHS), assesses the health and nutritional status of the U.S. population. Annually, NHANES examines about 5,000 nationally representative individuals through interviews and physical exams, collecting data on demographics, socioeconomic factors, diet, and health, along with medical measurements and lab tests.

We thoroughly reviewed NHANES data, focusing on health information from various reports. The diet section surveys the frequency of consumption of various foods, including soft drinks, alcohol, hamburgers, and salads. The examination section provides physical measurements like blood pressure and body weight, while the lab section includes biomarker assays such as blood sugar levels.

For the topic under discussion, our initial step involves sourcing data relevant to Diet Coke consumption and its association with fatness. Numerous studies~\citep{Adabk1274,BMI,BMI2}, have validated the Body Mass Index (BMI) as an optimal measure for assessing obesity, hence we will utilize BMI data as our indicator of obesity in this paper. However, regarding the intake of Diet Coke, due to changes in dietary survey questionnaires, relevant data was collected only during two survey cycles between 2003 and 2006. Nonetheless, the formulation of Diet Coke has not altered over the past twenty years~\citep{CocaColaFAQ2023}, making the use of this data both reasonable and valid for our analysis.

We hypothesize that the relationship between Diet Coke consumption and obesity is not straightforward but influenced by multiple factors. Thus, identifying key indicators impacting obesity is crucial. ~\citep{visser1997density} suggests that \textit{age, race, and gender} may affect obesity due to differences in body density. ~\citep{brown2016exploring} argues that \textit{education} can influence body fat through changes in savings behavior. ~\citep{akbartabartoori2005relationships} indicates a link between \textit{smoking} and obesity, moderated by waist circumference (WC), hip circumference (HC), and waist-to-hip ratio (WHR). ~\citep{elagizi2020review} states that \textit{physical activity} impacts weight by affecting the cardiovascular system. ~\citep{zhou2018body}'s analysis shows that within a normal BMI range, leaner individuals have a lower risk of \textit{hypertension}. ~\citep{carr2004abdominal} highlights that abdominal fat, an obesity marker, is closely associated with the risks of \textit{hyperlipidemia} and Type 2 \textit{diabetes} linked to metabolic syndrome.

Based on the aforementioned studies, the factors identified as influencing fatness include \textit{age, gender, race/ethnicity, educational level, smoking status, average daily physical activity level, hyperlipidemia, and diabetes mellitus.}

\subsection{Data Processing}

\begin{figure}
    \centering
    \includegraphics[width=1\linewidth]{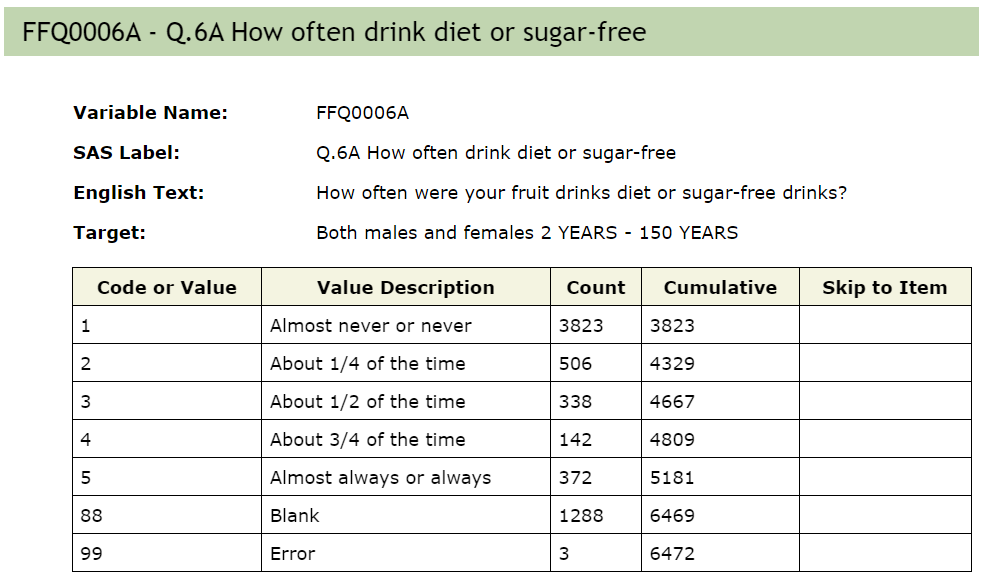}
    \caption{Diet or sugar-free Consumption}
    \label{causal2}
\end{figure}
\begin{figure}
    \centering
    \includegraphics[width=1\linewidth]{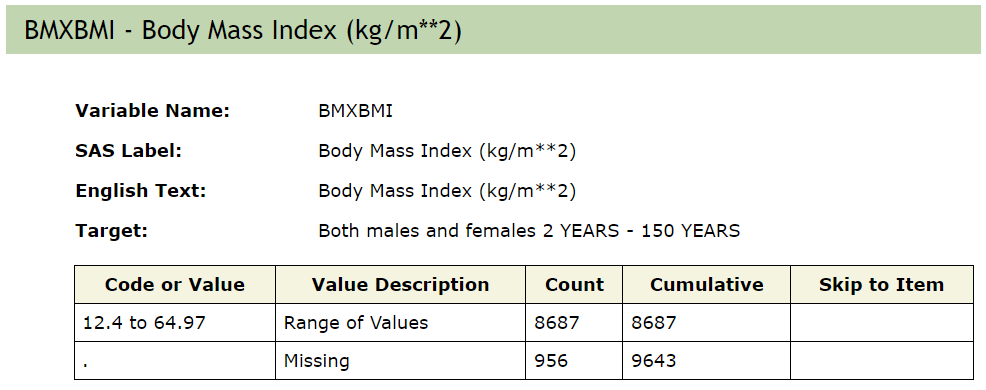}
    \caption{BMI measurements}
    \label{causal3}
\end{figure}
We extracted the necessary data for constructing the causal diagram and clarified the data types for each variable. As shown in Figure~\ref{causal2}, Diet Coke consumption frequency is categorized into five levels, coded from 1 to 5, ranging from ``Almost never or never" to ``Almost always or always," with corresponding respondent counts. Special cases like non-responses (``Blank") and data entry errors (``Error") are also noted.

In contrast, the fatness variable, represented by BMI, consists of continuous values providing precise measurements, as illustrated in Figure~\ref{causal3}. Other influencing factors are classified as either discrete or continuous. With this data, we can now construct the causal diagram.



To clearly define the relationships among variables and ensure methodological accuracy, we identify the independent variable ($X$), the dependent variable ($Y$), and relevant covariates:



The independent variable ($X$) is ``Diet Coke consumption frequency," quantifying how often participants consume Diet Coke.

The dependent variable ($Y$) is ``fatness," measured by BMI, calculated from participants' weight and height.

Additional covariates include age, gender, race/ethnicity, education, and health factors like smoking and physical activity. These covariates help us better understand the relationship between Diet Coke consumption and fatness.

\renewcommand{\arraystretch}{1.15}
\begin{table}[]
\centering
\caption{Summary of Variables}
\begin{tabular}{|c|l|}
\hline
\textbf{Category} & \textbf{Variable Description} \\ \hline
$X$ & Diet Coke Consumption Frequency \\ \hline
$Y$ & Fatness (BMI) \\ \hline
& Age \\ \cline{2-2} 
& Gender  \\ \cline{2-2} 
& Race / Ethnicity  \\ \cline{2-2} 
& Education Level  \\ \cline{2-2} 
& Smoking Status  \\ \cline{2-2} 
& Average Daily Physical Activity Level \\ \cline{2-2} 
& Hyperlipidemia  \\ \cline{2-2} 
\multirow{-9}{*}{Covariates} & Diabetes Mellitus  \\ \hline
\end{tabular}
\label{tab:variables}
\end{table}

For the measurement of various health indicators, authoritative standards are essential. According to ~\citep{CDC2022BMI}, a Body Mass Index (BMI) of 30.0 or higher categorizes an individual within the obesity range. Hyperlipidemia, or high cholesterol, is characterized by certain thresholds in cholesterol measurements~\citep{MayoClinic2024}. It is considered high when total cholesterol exceeds 240 mg/dL, LDL (low-density lipoprotein) cholesterol is 160 mg/dL or higher, and triglycerides are above 200 mg/dL. HDL (high-density lipoprotein) cholesterol is ideally above 60 mg/dL, as lower levels may also indicate risk. The A1C test measures the percentage of glycated hemoglobin, providing an average blood glucose level over three months. Levels of $5.7\%$ to $6.4\%$ indicate prediabetes, while $6.5\%$ or higher confirms diabetes.

\subsection{Causal Diagram}
In the field of causal inference, the $IC^{\ast}$ algorithm is a pivotal tool for identifying causal relationships among variables from observational data~\citep{pearl2009}. This study employs the $IC^{\ast}$ algorithm to construct a causal graph, delineating the causal structures between variables $X$,$Y$, and several covariates. The initial step of the $IC^{\ast}$ algorithm involves conducting conditional independence tests to analyze relationships among all pairs of variables within the dataset. This process relies on statistical tests to evaluate whether different combinations of variables are independent; if two variables are found to be dependent given other variables, an undirected edge is established between them.

Subsequently, the algorithm enters the structure identification phase, utilizing conditional independence tests, specifically \textit{the Robust Regression Test}, to detect $``V-structures''$ within the data.
These structures are unique triplets where one variable is a common cause of the other two variables, which do not directly interact with each other. Identifying these V-structures is crucial for determining the direction of the edges in the graph, as they reveal the causal pathways, either direct or through intermediary variables.

\begin{figure}
    \centering
    \includegraphics[width=0.95\linewidth]{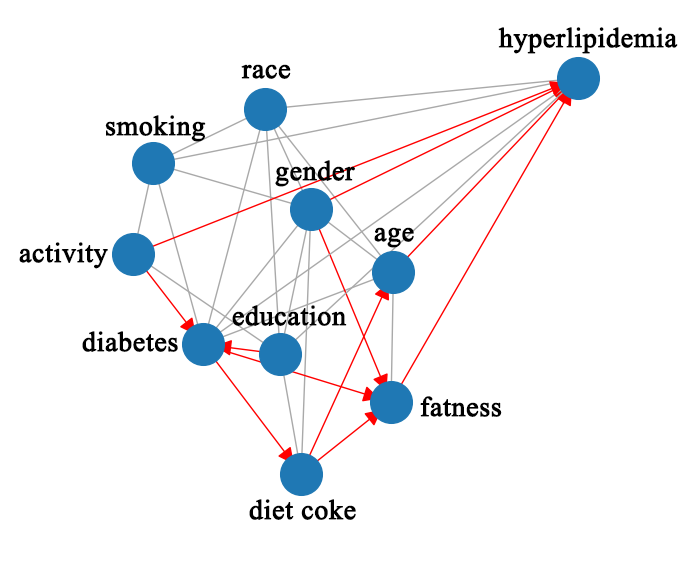}
    \caption{Causal Diagram($IC^{\ast}$ algorithm)}
    \label{causal4}
\end{figure}

Furthermore, the $IC^{\ast}$ algorithm is capable of identifying latent variables within the causal graph and refining the graph structure through iterative optimization processes. In each iteration, the algorithm evaluates whether adding or adjusting the direction of edges better conforms to the evidence of conditional independence from the data. This method allows the algorithm to gradually construct a detailed and accurate map of the causal relationships between the variables. By applying the $IC^{\ast}$ algorithm, we successfully portray the complex causal network among variables $X$, $Y$, and the covariates, providing a solid foundation for further analysis and model construction.The resulting causal graph is depicted in Figure~\ref{causal4}. Red arrows denote clear causal chains, while solid black lines indicate potential but uncertain connections. 


To ensure the accuracy of this causal diagram, we used the reliable DAG with NO TEARS algorithm~\citep{zheng2018dags}. By comparing the diagrams generated by both methods, we found them nearly identical, further validating the correctness of our causal graph from the $IC^{\ast}$ algorithm.

\subsection{RCT Calculation}
In this section, we focus on calculating counterfactual values~\citep{pearl2016causal}, which represent potential outcomes under the theoretical full control of a specific variable, such as the frequency of Diet Coke consumption. Counterfactual analysis is a central concept in causal inference, enabling us to explore the potential relationships between variables under various interventional scenarios. This means assessing the outcomes when one factor is altered while all others are held constant.

For variable $X$, representing the frequency of Diet Coke consumption, the original data categorize consumption into five levels ranging from 1 to 5. For computational convenience, we binarized these data, designating those who never drink Diet Coke as 0, and marking all other levels as 1. For $Y$ (fatness), individuals are classified as 0 for non-obese and 1 for obese, following the criteria previously described.

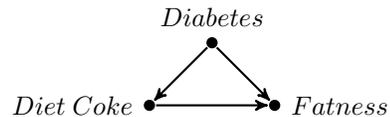
\begin{figure}[h]
\centering
\begin{tikzpicture}[->,>=stealth',node distance=1.2cm,
  thick,main node/.style={circle,fill,inner sep=1.5pt}]
  \node[main node] (1) [label=above:{$Diabetes$}]{};
  \node[main node] (2) [below left =1cm of 1,label=left:$Diet\ Coke$]{};
  \node[main node] (3) [below right =1cm of 1,label=right:$Fatness$] {};
  \path[every node/.style={font=\sffamily\small}]
    (1) edge node {} (2)
    (1) edge node {} (3)
    (2) edge node {} (3);
\end{tikzpicture}
\caption{The segment of the causal diagram that satisfies the backdoor criterion.}
\label{causal6}
\hfill

\end{figure}

Upon analyzing the causal diagram derived earlier, we identified a single backdoor pathway as shown in Figure~\ref{causal6}, which is characterized by the pathway $Diet Coke\rightarrow{}Diabetes\leftarrow{}fatness$. This aligns with the findings of ~\citep{gardener2018diet}, which demonstrated a significant association between Diet Coke consumption and an increased risk of diabetes. Consequently, controlling for diabetes, the only confounder in this pathway, allows us to use the adjustment formula~\citep{pearl1995} to compute the experimental data. This calculation estimates the causal impact of Diet Coke consumption on fatness, controlled for diabetes status. These are denoted as $P(Y=1 \mid \text{do}(X=1))$
 and $P(Y=1 \mid \text{do}(X=0))$, also represented in our notation as $y_{x}$ and $y_{x'}$.

The challenge of discerning the actual effectiveness in these sensitive contexts stems from their counterfactual\footnote{Counterfactuals are statements or scenarios that express what might have happened if circumstances had been different from what actually occurred.} nature; outcomes of interest are inferred through hypothetical scenarios rather than directly observed. For example, if an individual consumes Diet Coke, the outcome had they not consumed it will never be observable.

RCT may seem sufficient to handle this counterfactual nature, but they fall short because no two individuals are exactly the same. Individuals may belong to subpopulations unknown to the investigator, each responding to a given treatment markedly differently~\cite{li2022unitselectioncasestudy}. Therefore, a more sophisticated approach is required to accommodate individual differences.

 \subsection{PNS Calculation}
First of all, We define the following variables and their states:

$X$: Represents the consumption status of Diet Coke. This is a binary variable where:
    \begin{itemize}
        \item  \(x = 1\) indicates that Diet Coke is consumed.
        \item  \(x' = 0\) or \(x'\) indicates that Diet Coke is not consumed.
    \end{itemize}

$Y$: Represents the obesity status of an individual. This is also a binary variable where:
  \begin{itemize}
      \item  \(y = 1\) indicates that the individual is obese.
      \item  \(y' = 0\) or \(y'\) indicates that the individual is not obese.
  \end{itemize}

With the basic variables and states defined, we use counterfactual notation to explore hypothetical changes in these states and their impact on outcomes:

$Y_x = y$: This expression is used to describe a scenario where if \(X\) were set to \(x\) (consuming Diet Coke), then \(Y\) would take the value \(y\) (obese).

$Y_{x'} = y$: This expression describes another scenario where, even without consuming Diet Coke (\(X=x'\)), the individual still reaches an obese state (\(Y=y\)).

Building on the foundation laid by the clear definitions and the use of counterfactual notation, the computation of the Probability of Necessity and Sufficiency (PNS) in our study is pivotal for understanding the causal impact of Diet Coke consumption on obesity. 

PNS evaluates the likelihood that consuming Diet Coke is both necessary and sufficient for the occurrence of obesity within various demographics. \textbf{Specifically, it measures the probability that obesity would occur with Diet Coke consumption and would not otherwise. } By calculating PNS, we aim to determine the extent to which Diet Coke consumption can be considered a direct causal factor in obesity.

With the experimental data estimated, we are now positioned to apply the model developed by \citep{TianPearl2000} to compute the bounds of the Probability of Necessity and Sufficiency ($PNS$) across the general population. Furthermore, employing the method outlined by ~\citep{ijcai2022p376}, we will refine these estimates by narrowing the bounds, as elaborated in Formulas (3)(4)(7)(8) of the $Preliminaries$ section of this paper.


Building on this foundational analysis, we segment the population into various subgroups based on covariates related to demographic characteristics and dietary habits to compute the $PNS$ within these subgroups. Each subgroup is formed with at least 385 individuals to ensure a margin of error of no more than 0.05 for the $95\%$ confidence interval~\citep{li2022probabilities}.

This stratification is essential as different dietary habits can bias outcomes. For instance, those who frequently consume beer, burgers, and doughnuts may have a higher risk of obesity, which shouldn’t be solely linked to Diet Coke. While Diet Coke has no sugar, it may still contribute to obesity by stimulating hormonal secretions or misleading brain mechanisms.

By carefully considering these factors, our aim is to identify specific demographics where Diet Coke consumption significantly leads to obesity (indicating a high lower bound of PNS) or demographics where Diet Coke has virtually no impact on obesity levels (reflecting a low upper bound of PNS). This nuanced approach allows us to deliver a more precise understanding of the causal dynamics between Diet Coke consumption and obesity across different subpopulations.

\section{Results}


A/B testing, like Randomized Controlled Trials (RCTs), is the gold standard for establishing causality in clinical and behavioral research. RCTs provide a controlled environment to directly compare and demonstrate the impact of interventions, such as Diet Coke consumption. Therefore, we initially estimated the counterfactual data for the general population, providing a clear and statistically robust basis for assessing the effects of Diet Coke on obesity prevalence. 
\begin{eqnarray*}
P(Y=1 \mid \text{do}(X=1)) = 0.4157292166, \\
P(Y=1 \mid \text{do}(X=0)) = 0.3198820277.
\end{eqnarray*}

Based on the analysis of simulated RCT data, we observed a significant correlation between the consumption of Diet Coke and obesity. Specifically, the probability of obesity among participants who consumed Diet Coke was $41.57\%$, compared to $31.99\%$ among those who did not consume Diet Coke. This indicates that the consumption of Diet Coke can increase the likelihood of obesity by approximately $9.58\%$. While this data suggests that Diet Coke may be a potential risk factor for obesity, the modest increase of less than $10\%$ across the general population may not be entirely convincing in complex real-world scenarios. In this context, the method of calculating the Probability of Necessity and Sufficiency (PNS) boundaries used in this study offers a more detailed and comprehensive analysis, revealing further nuances and complexities associated with Diet Coke's impact on obesity. 

Following the methods described previously, we computed and refined the boundaries of the Probability of Necessity and Sufficiency (PNS) for the general population:
\begin{eqnarray*}
0.096 \le PNS \le 0.405.
\end{eqnarray*}
The calculated Probability of Necessity and Sufficiency (PNS) for obesity attributable to Diet Coke consumption ranges from 0.096 to 0.405. The lower bound of approximately 10$\%$ suggests that there indeed exists a subset of individuals for whom consuming Diet Coke contributes to obesity. However, the broad range of this boundary indicates variability in the causal impact across the population. 

To refine our understanding and identify potentially more characteristic subgroups, we will next calculate the PNS boundaries for various subpopulations. This approach aims to discover specific demographics or behavioral patterns that might exhibit a stronger or more distinct causal relationship between Diet Coke consumption and obesity.

We have selected some subpopulations where the effects are particularly pronounced, as shown in Table 2. After controlling for certain covariates, the PNS boundaries underwent significant changes. For instance, in the subgroup labeled as $Old\_Man\_Activity\_Low$, which refers to men over the age of 60 with low physical activity levels, the lower boundary of PNS increased by more than double compared to the general population. This indicates that the impact of Diet Coke on weight is more significant in this subgroup.

As shown in Table 3, we further subdivided the population based on dietary habits, yielding significant results. Among individuals with poor dietary habits, such as the subgroup $``Old\_Man\_Hamburger"$, which includes elderly males who frequently consume hamburgers, the lower boundary of PNS rose to nearly $30\%$.This suggests that a higher proportion of individuals in this group experience obesity potentially linked to consuming Diet Coke, illustrating the significant probability that obesity would occur with Diet Coke consumption and would not otherwise. However, it is likely not the consumption of Diet Coke itself that leads to obesity, but rather the association of Diet Coke with a range of unhealthy dietary behaviors.

Conversely, in subgroups with healthier eating habits, such as $``Young\_Woman\_No\_Hotdog"$, the upper boundary being $20\%$ indicates that only a small fraction of this population might become obese due to Diet Coke consumption. In these cases, the practice of consuming Diet Coke while maintaining weight stability appears to be more credible. This differential impact underscores the complex interactions between dietary habits and the effects of specific dietary choices such as Diet Coke on health outcomes.

\renewcommand{\arraystretch}{1.2} %

\begin{table}[t]

\centering 
\caption{PNS by Subpopulation Based on Demographics}
\begin{tabular}{|l|l|}
\hline

\textbf{Subpopulation}                    & \textbf{Bounds of PNS}  \\ \hline
Activity\_High                     & 0.10122, 0.39544            \\ \hline
Man                              & 0.12298, 0.40726            \\ \hline
Age60+                           & 0.14792, 0.43166            \\ \hline
Old\_Man                      & 0.16865, 0.41567            \\ \hline
Old\_Activity\_Low              & 0.14637, 0.47675            \\ \hline
Old\_Hyperlipidemia\_Yes        & 0.15442, 0.40864            \\ \hline
Old\_Man\_Education\_Low              & 0.16102, 0.41829            \\ \hline
Old\_Man\_Activity\_Low         & 0.22938, 0.53263            \\ \hline
Woman                            & 0.08430, 0.39075            \\ \hline
Age60-                           & 0.09140, 0.37639            \\ \hline
Young\_Woman                    & 0.07391, 0.37845            \\ \hline
Young\_Woman\_Activity\_High      & 0.05077, 0.34600            \\ \hline
\end{tabular}

\end{table}

\begin{table}[t]
\centering 
\caption{PNS by Subpopulation Based on Dietary Habits}
\begin{tabular}{|l|l|}
\hline
\textbf{Subpopulation }                              & \textbf{Bounds of PNS }   \\ \hline
Old\_Man\_Hamburger & 0.29909, 0.57860 \\ \hline
Old\_Man\_Hotdog                            & 0.24406, 0.48335 \\ \hline
Old\_Man\_Fries                             & 0.20028, 0.47119 \\ 
\hline
Old\_Man\_Icecream                    & 0.18539, 0.47663 \\ 
\hline
Old\_Man\_Candy                              & 0.16524, 0.37878 \\ \hline
Old\_Man\_Beer             & 0.14397, 0.42206 \\ \hline
Young\_Woman\_No\_Hamburger                 & 0.00905, 0.22395 \\ \hline
Young\_Woman\_No\_Popcorn                   & 0.02081, 0.29678 \\ \hline
Young\_woman\_Salad                         & 0.07663, 0.29593 \\ \hline
Young\_Woman\_No\_Syrup                   & 0.01923, 0.23076 \\ \hline
Young\_Woman\_No\_Fries                     & 0.00198, 0.21941 \\ \hline
Young\_Woman\_No\_Hotdog                    & 0.00000, 0.19487 \\ \hline
\end{tabular}

\end{table}

\section{Discussion}

In this paper, we employed Structured Causal Models (SCM) and the $IC^{\ast}$ algorithm to construct causal diagrams and calculate the probability of interventional effects using adjustment formulas. We then assessed the likelihood of obesity with and without Diet Coke consumption. However, our study has some limitations.

One key limitation is the constrained sample size, which required each subgroup to have at least 385 individuals to ensure reliable statistical results~\citep{li2022probabilities}. This constraint limited our ability to perform more detailed subgroup analyses, making it difficult to precisely identify which populations are more or less affected by Diet Coke consumption. Future research should expand the sample size and include a broader range of regions and populations to improve the generalizability and accuracy of the findings.


Additionally, our study may have relied on self-reported data, particularly regarding dietary habits, which can be prone to bias. For instance, a participant's recent increase in Diet Coke consumption might lead to overestimations of their usual intake, introducing inaccuracies. Incorporating more objective measures, such as biomarkers or direct observations, in future studies could enhance data reliability and research validity.

Furthermore, although studies~\citep{sylvetsky2020consumption} suggest that Diet Coke may influence obesity through hormone secretion, we lacked specific hormone data, limiting our understanding of these potential mechanisms. Future research should explore the effects of Diet Coke on hormones like insulin and leptin to clarify its relationship with obesity and metabolic health. Additionally, examining the connections between Diet Coke consumption and other chronic diseases, such as cardiovascular disease and diabetes, would provide a more comprehensive view of its health implications.

\section{Conclusion}
Through the application of causal inference methods, our study has determined that Diet Coke consumption indeed poses a risk of increased obesity; however, the impact varies significantly across different demographics. Different subgroups experience varying rates of weight gain after consuming Diet Coke, suggesting that the influence of Diet Coke on weight is not a straightforward causal relationship but is likely mediated through multiple factors. These findings emphasize the importance of considering individual differences and dietary habits when assessing the health impacts of foods or beverages. This complexity highlights that the effect of Diet Coke on weight gain is the result of multifaceted interactions, underscoring the necessity to take into account personal health profiles and lifestyles in dietary recommendations.

\bibliography{aaai25}
\clearpage
\section{Reproducibility Checklist}

\subsection*{This paper:}
\begin{itemize}
    \item Includes a conceptual outline and/or pseudocode description of AI methods introduced: \textbf{yes}
    \item Clearly delineates statements that are opinions, hypotheses, and speculation from objective facts and results: \textbf{yes}
    \item Provides well-marked pedagogical references for less-familiar readers to gain background necessary to replicate the paper: \textbf{yes}
    \end{itemize}

\subsection*{Does this paper make theoretical contributions? \textbf{No}}

\subsection*{Does this paper rely on one or more datasets? \textbf{yes}}

\begin{itemize}
    \item A motivation is given for why the experiments are conducted on the selected datasets: \textbf{yes}
    \item All novel datasets introduced in this paper are included in a data appendix: \textbf{yes}
    \item All novel datasets introduced in this paper will be made publicly available upon publication of the paper with a license that allows free usage for research purposes: \textbf{yes}
    \item All datasets drawn from the existing literature (potentially including authors’ own previously published work) are accompanied by appropriate citations: \textbf{yes}
    \item All datasets drawn from the existing literature (potentially including authors’ own previously published work) are publicly available: \textbf{yes}
    \item All datasets that are not publicly available are described in detail, with an explanation of why publicly available alternatives are not scientifically satisfying: \textbf{NA}
\end{itemize}

\subsection*{Does this paper include computational experiments? \textbf{No}}

\end{document}